\documentclass[twocolumn,aps,prb,superscriptaddress
]{revtex4-1}

\usepackage{times}
\usepackage{graphicx}
\usepackage{amsfonts}
\usepackage{amsmath, amsthm, amssymb}
\usepackage{dsfont}
\usepackage{mathptm}
\usepackage{color}
\usepackage{standalone}
\usepackage{tikz}
\usetikzlibrary{decorations.markings}
\usetikzlibrary{shapes,positioning,arrows}
\usepackage{bm}
\usepackage[colorlinks=true, allcolors=black, citecolor=blue, linkcolor=blue]{hyperref}

\renewcommand{\rm}[1]{\mathrm{#1}}
\newcommand{\su}{\uparrow}
\newcommand{\sd}{\downarrow}

\newcommand{\s}{\sigma}
\newcommand{\barm}{\bar{m}}

\renewcommand{\L}{\mathcal{L}}
\newcommand{\D}{\mathcal{D}}

\newcommand{\rmd}{\mathrm{d}}

\newcommand{\del}{\partial}

\newcommand{\vk}{{\bf{k}}}
\newcommand{\vq}{{\bf{q}}}

\newcommand{\vm}{{\bf{m}}}
\newcommand{\tvm}{\tilde{\bf{m}}}
\newcommand{\tm}{\tilde{m}}
\renewcommand{\vr}{{\bf{r}}}
\newcommand{\vf}{v_\rm{F}}
\newcommand{\kf}{k_\rm{F}}

\renewcommand{\v}[1]{{\bf{#1}}}
\newcommand{\vs}{\bm{\s}}

\newcommand{\psiu}{\psi_\su}
\newcommand{\psid}{\psi_\sd}

\renewcommand{\O}[1]{\mathcal{O}(#1)}

\newcommand{\etal}{{\emph{et al.}}~}
\newcommand{\ie}{{i.e.},~}
\newcommand{\eg}{{e.g.}~}

\newcommand{\J}{\bar{J}}

\usepackage{soul}

\allowdisplaybreaks[1]

\begin{document}

\title{Magnon-induced superconductivity in a topological insulator coupled to ferro- and antiferromagnetic insulators}

\author{Henning G. Hugdal}
\affiliation{Department of Physics, NTNU, Norwegian University of Science and Technology, NO-7491 Trondheim, Norway}
\affiliation{Center for Quantum Spintronics, Norwegian University of Science and Technology, NO-7491 Trondheim, Norway}

\author{Stefan Rex}
\affiliation{Institut f\"ur Nanotechnologie, Karlsruhe Institute of Technology, 76021 Karlsruhe, Germany}
\affiliation{Institut f\"ur Theorie der Kondensierten Materie, Karlsruhe Institute of Technology, 76128 Karlsruhe, Germany}

\author{Flavio S. Nogueira}
\affiliation{Institute for Theoretical Solid State Physics, IFW Dresden, PF 270116, D-01171 Dresden, Germany
}

\author{Asle Sudb{\o}}
\email[Corresponding author: ]{asle.sudbo@ntnu.no}
\affiliation{Department of Physics, NTNU, Norwegian University of Science and Technology, NO-7491 Trondheim, Norway}
\affiliation{Center for Quantum Spintronics, Norwegian University of Science and Technology, NO-7491 Trondheim, Norway}


\begin{abstract}
We study the effective interactions between Dirac fermions on the surface of a three-dimensional topological insulator due to the proximity coupling to the magnetic fluctuations in a ferromagnetic or antiferromagnetic insulator. Our results show that the magnetic fluctuations can mediate attractive interactions between Dirac fermions of both Amperean and BCS type. In the ferromagnetic case, we find pairing between fermions with parallel momenta, so-called Amperean pairing, whenever the effective Lagrangian for the magnetic fluctuations does not contain a quadratic term. The pairing interaction also increases with increasing Fermi momentum, and is in agreement with previous studies in the limit of high chemical potential. If a quadratic term is present, the pairing is instead of BCS type above a certain chemical potential. In the antiferromagnetic case, BCS pairing occurs when the ferromagnetic coupling between magnons on the same sublattice exceeds the antiferromagnetic coupling between magnons on different sublattices. Outside this region in parameter space, we again find that Amperean pairing is realized.
\end{abstract}

\maketitle

\section{Introduction}
Topological insulators (TIs) have attracted much attention since their discovery a decade ago.\cite{Hasan2010,Qi2011} Though being insulating in the bulk, the surface of a three-dimensional (3D) TI has topologically protected, metallic surface states. These metallic surface states are well described by the two-dimensional (2D) massless Dirac equation, having linear dispersions and spin-momentum locking, and are therefore often called Dirac fermions.\cite{Wehling2014} A gap in the dispersion, analogous to the mass gap for massive relativistic fermions, can be opened by breaking the time-reversal symmetry of the system, for instance by applying a magnetic field normal to the TI surface, or by proximity coupling to a magnetic insulator.\cite{Wei2013,Yang2013,Liu2015,Katmis2016,Tang2017} 

Many theoretical works have studied heterostructures consisting of TIs and ferromagnetic (FM) insulators, focusing in particular on the effects on the magnetization in the magnetic layer.\cite{Garate2010,Nomura2010,Tserkovnyak2012,Semenov2012,Ferreiros2014,Linder2014,Semenov2014,Ferreiros2015,Duan2015,Rex2016,Rex2016a,Rex2017} One recent study focused instead on the effective interactions between Dirac fermions on the surface of a TI coupled to a FM insulator with mean field magnetization perpendicular to the TI surface.\cite{Kargarian2016} It showed that interactions between the Dirac fermions and the transverse magnons in the FM lead to an effective attractive pairing between fermions with parallel momenta, so-called Amperean pairing.\cite{Lee2007,Lee2014} In the presence of spin-momentum locking, this exotic pairing also implies that the pairs will form spin triplets. However, the chemical potential was assumed to be tuned far away from the gap, thus neglecting the effects of the mass gap in the Dirac fermion dispersion. This raises the question how the pairing is affected when the chemical potential is tuned towards the gap, as the pairing must disappear in the absence of a Fermi surface. Moreover, the pairing mediated by fluctuations in other magnetic configurations than FM order have not yet been studied. Bilayer systems of antiferromagnetic (AFM) insulators and TI films, for instance, are also under experimental investigation\cite{He2017}.

Rex \etal\cite{Rex2017} recently studied the effective theory for the magnetic moments in a bipartite magnetic insulator (BMI) coupled to the Dirac fermions on a TI surface. Their model allows to continously tune the magnet from an FM to an AFM configuration.
In the present paper we will use the same model, restricted to the limiting FM and AFM cases, to study the effective interactions between the Dirac fermions induced by the magnetic fluctuations, including the effects of the mass gap. Possible material choices for such systems are Bi$_2$Se$_3$ or Bi$_2$Te$_3$ as TI, EuS as FM,\cite{Wei2013,Li2015} and NiO or CoO as AFM.\cite{Wang2014,Hahn2014,Moriyama2015,Wang2015,Lin2016} In both cases, we find that pairing between Dirac fermions is possible in certain regions of parameter space. For coupling to ferromagnetic fluctuations, the pairing is of the Amperean type {whenever there is no quadratic coupling term between the magnons}, in agreement with Ref.~\onlinecite{Kargarian2016} in the limit of high chemical potential. However, as the Fermi level is moved towards the gap, the pairing decreases, vanishing when the chemical potential is tuned inside the gap. {We also find that pairing of the Bardeen-Cooper-Schrieffer (BCS) type, \ie where the interacting particles have momenta in opposite directions, is possible in certain regions of parameter space. In the antiferromagnetic case we again find pairing of both types, depending on the relative strenght between the intra- and interlattice coupling.} Hence, we find that for both magnetic configurations, magnon-induced superconductivity due to the attractive interactions is possible.

The remainder of the article is organized as follows: The model is presented in Sec.~\ref{sec:model}, together with the derivation of the effective action for the TI surface fermions. Subsequently, the effective pairing interaction on the TI surface is discussed for the FM and AFM cases in Sec.~\ref{sec:FM} and \ref{sec:AFM}, resepectively. The results are summarized in Sec.~\ref{sec:conclusion}. Further details regarding the derivation of the effective TI action are presented in the Appendix.

\section{Model}\label{sec:model}
\begin{figure}
\includegraphics[width=0.85\columnwidth]{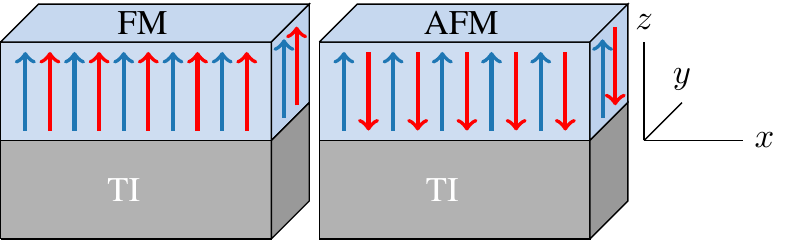}
\caption{\label{fig:model}Bilayer heterostructures consisting of ferromagnetic and antiferromagnetic insulators proximity coupled to a TI are modelled using a bipartite magnetic insulator with tunable mean field magnetizations.\cite{Rex2017} The surface of the TI is placed in the $xy$-plane, and the mean field magnetization of the magnetic insulators is perpendicular to the interface.}
\end{figure}

The bilayer heterostructures are described by taking into account the surface of the TI and magnetic insulator, the bulk of the magnetic insulators, and hopping across the interface due to the proximity.\cite{Rex2017} In order to treat heterostructures with FM and AFM insulators simultaneously, we will consider a BMI consisting of two FMs with lattice magnetizations $\vm_1$ and $\vm_2$, as is illustrated in Fig.~\ref{fig:model}.
We set $\hbar =1$ throughout the paper, and work close to zero temperature, utilizing the zero-temperature Matsubara frequency formalism. The bulk of the magnetic insulator is described by the Lagrangian $\L_\rm{BMI} = \L_1 + \L_2 + \L_{\rm{ex}}$, where
\begin{equation}
    \L_i = -\v{b}(\vm_i) \cdot \del_t \vm_i - \frac{\kappa}{2}(\nabla \vm_i)^2
\end{equation}
amounts to a continuum description of each of the two sublattices with $i=1,2$, while
\begin{equation}
    \L_\rm{ex} = -\lambda \vm_1\cdot \vm_2
\end{equation}
describes the exchange interaction. $\kappa >0$ is the ferromagnetic exchange coupling constant, while the coupling between the two lattices is ferromagnetic or antiferromagnetic for $\lambda <0$ or $\lambda >0$, respectively. $\v{b}$ denotes the Berry connection, satisfying $\nabla_{\vm_i} \times \v{b}(\vm_i) = \vm_i/m_i^2$.

The surface of the TI is described by the 2D Dirac Lagrangian, together with a weak quadratic term in the derivatives leading to particle-hole asymmetry,\cite{Taskin2011,Wright2013,Li2013,Leblanc2014}
\begin{equation}
    \L_\rm{TI} = \Psi^\dagger\bigl[i\del_t - i\vf (\s_y \del_x - \s_x \del_y) + E_0(\del_x^2+\del_y^2)+ \mu\bigr]\Psi,
\end{equation}
where $\Psi = (\psi_{\su},~ \psi_{\sd})^T$ is the spinor of the Dirac fermions, $\uparrow, \downarrow$ label the spin in z direction, $\vf$ is the Fermi velocity, and $\mu$ is the chemical potential. The second derivative term is assumed small compared to the Dirac term. We have not included the fluctuating Coulomb interactions between the Dirac fermions, since this interaction is screened whenever we have a Fermi surface. For the ferromagnetic case, there will also be a demagnetizing field outside the ferromagnet, resulting in a coupling to a mostly in-plane vector potential in the TI Hamiltonian.\cite{Griffiths2013} This coupling in turn leads to circular orbits with radius of the order of the magnetic length $l \sim \sqrt{\hbar/e\mu_0|\v{M}|}$\cite{Li2013} where $\v{M}$ is the magnetization, $\mu_0$ the vaccum permeability, and $e$ the electron charge. This coupling can only be neglected when the motion of the TI fermions are unaffected on the relevant length scale, which for superconductivity is the coherence length $\xi$, \ie we must have $l \gg \xi$. Using $\xi \sim \hbar\vf/k_\mathrm{B} T_c$\cite{Bruus2004} where $T_c$ is the critical temperature, we get the requirement that $|\v{M}|\ll k_\mathrm{B}^2 T_c^2/e\mu_0\hbar \vf^2.$ We will assume that this holds in the following. Since antiferromagnets have close to zero stray fields,\cite{Marti2014,Marti2015} the coupling to the vector potential can be safely disregarded in the AFM case.

In order to couple the TI fermions and BMI magnetization, Rex \etal \cite{Rex2017} introduced auxiliary fermionic fields $\chi_i = (\chi_\su, \chi_\sd)^T$ on the surface of the magnet for the two sublattices, $i=1,2$. These fields can be interpreted as electrons in the magnetic insulator, which are localized in the atomic limit. Their spins $\v{S}_i=\frac12\chi_i^\dagger \boldsymbol{\sigma}\chi_i$ are coupled to the magnetization of the corresponding sublattice, and in proximity to the TI hopping across the interface is taken into account. Thus, the Hamiltonian of $\chi_1, \chi_2$ is
\begin{equation}\label{H_surf}
\begin{aligned}
    H_\rm{surf} ={} &-t(\chi_1^\dagger\chi_2 + \chi_2^\dagger\chi_1) - J\sum_{i=1,2}\chi_i^\dagger\vm_i\cdot\boldsymbol{\sigma}\chi_i \\
    & - h\left[\Psi^\dagger(\chi_1+\chi_2) + (\chi_1^\dagger + \chi_2^\dagger)\Psi\right].
\end{aligned}
\end{equation}
Here, $t$ is the coupling between the surface fermions, $J$ and $h$ are the strengths of the coupling to the magnetization $\vm_i$ at $z=0$ and to the Dirac fermions respectively, and $\vs$ is the Pauli matrix vector.

\subsection{Integration of magnetic moments}
By integrating out the $\chi_i$ fermions, an effective theory for the Dirac fermions and magnetizations was obtained in Ref.~\onlinecite{Rex2017}, including effective couplings between $\Psi$ and $\vm_i$. In the following, we will assume that the length of the magnetizations $\vm_i$ is fixed to the mean field value, $|\vm_i|=\pm\barm_i$, and write the magnetization vector as\cite{Kargarian2016}
\begin{equation}
    \vm_i = \barm_i\hat{z}\sqrt{1-\frac{\tvm_i^2}{\barm_i^2}}+\tvm_i,
\end{equation}
where $\tvm_i = \tm_i^x\hat{x} + \tm_i^y\hat{y}$. By fixing the length in this way, the fluctuations in the $z$-direction are of second order in $|\tvm_i|$. Working to second order in $\tm_i^{x/y}$, we get the Berry connection
\begin{equation}
    \v{b}(\vm_i) = -\frac{\tm_i^y\hat{x} - \tm_i^x\hat{y}}{2\barm_i},
\end{equation}
and the effective Lagrangian for the magnetic fluctuations can be written as
\begin{equation}
\label{L_mag_eff}
\begin{aligned}
    \L_\rm{\tvm} &= \sum_{i=1,2} \Bigg\{-\left(\frac{1}{2\barm_i} -2\D_i^{0z}\right)\hat{z}\cdot\left(\tvm_i\times\del_t\tvm_i\right) - \frac{\kappa}{2}(\nabla\tvm_i)^2\\
    & + \frac{1}{2}\left[\frac{\barm_{3-i}}{\barm_i}\lambda - 2J^2\left(D_i^{00} - D_i^{zz}\right)\right]\tvm_i\cdot\tvm_i + \Psi^\dagger J_i\tvm_i\cdot\vs\Psi\Bigg\} \\
    & - \left[\lambda + 2J^2\left(T^{00} - T^{zz}\right)\right]\tvm_1\cdot\tvm_2  \\
    & + 2J^2\mathcal{T}^{0z} \hat{z}\cdot\big(\tvm_1\times\del_t\tvm_2 + \tvm_2\times\del_t\tvm_1\big),
\end{aligned}
\end{equation}
where
\begin{equation}
    J_i = \frac{h^2J}{(t^2-J^2\barm_1\barm_2)^2}(J^2 \barm_{3-i}^2-t^2),\label{J_i}
\end{equation}
is the effective magnetic coupling of the TI surface states to $\v{m}_i$. The coefficients $D_i^{00}$, $D_i^{zz}$, $\mathcal{D}_i^{0z}$, $T^{00}$, $T^{zz}$ and $\mathcal{T}^{0z}$ depend only on the model parameters and are described in detail in Ref.~\onlinecite{Rex2017}. Since $h$ is assumed small compared to $t$ and $J\barm_i$,\cite{Rex2017} we have neglected terms of $\O{h^2|\tvm_i|^2}$ in Eq.~(\ref{L_mag_eff}). Notice that the exchange couplings between fluctuations are renormalized in the above Lagrangian,
\begin{subequations}
\label{ex_coupling_fluctuations}
\begin{align}
    \tilde{\lambda}_i &= \frac{\barm_{3-i}}{2\barm_i} \lambda - J^2(D_i^{00} - D_i^{zz}),\\
    \lambda_\rm{eff} & = \lambda + 2J^2(T^{00} - T^{zz}).
 \end{align}    
\end{subequations}
Integration of the $\chi$ fermions also results in an additional term in the Dirac Lagrangian due to the mean field magnetizations in the BMI,
\begin{equation}
    \delta\L_\rm{MF} = \Psi^\dagger(J_1\barm_1 \s_z + J_2\barm_2\s_z)\Psi.\label{L_MF}
\end{equation}
As will be shown in the next section, this term can create a gap in the Dirac fermion dispersion.

Specializing to the ferromagnetic ($\barm_2 = \barm_1 = \barm$) and antiferromagnetic ($\barm_2 = -\barm_1 = -\barm$) cases, we define $\nu = \barm_2/\barm_1=\pm1$ for notational simplicity. In both cases the magnetic couplings in Eq.~(\ref{J_i}) are identical on the two sublattices, $J_1 = J_2 \equiv \J$. Transforming to imaginary time, $T = it$, in the zero temperature limit, and Fourier transforming both the time and space variables,\footnote{We have used the sign convention
\begin{equation}
    f(\vk,\omega) = \int \rmd T \int \rmd^2 r ~ e^{-i\vk\cdot\vr - i\omega T} f(T,\vr)
\end{equation}
when Fourier transforming.} we arrive at the functional integral in the magnon fields
\begin{equation}
    Z = \int \D[M] e^{-S_\rm{mag}},
\end{equation}
where
\begin{equation}
\begin{aligned}
    S_\text{mag} &= \int \frac{\rmd^3 q}{(2\pi)^3} \Big\{M^T(-q)\mathcal{K}(q)M(q) \\
    &\qquad- \frac{1}{2}\left[\mathcal{J}^T(q)M(q) + M^T(-q)\mathcal{J}(-q)\right]\Big\}.
\end{aligned}
\end{equation}
Here, we have defined the matrix
\begin{widetext}
\begin{equation}
    \mathcal{K}(q) = \left(\begin{matrix}
        \frac{\kappa}{2}\vq^2 - \frac{\lambda\nu}{2} + J^2 D - \frac{i\s_y}{2m^*}\Omega & \frac{\lambda}{2} + J^2 T + J^2(1+\nu)\mathcal{T}^{0z}i\s_y \Omega\\
        \frac{\lambda}{2} + J^2 T + J^2(1+\nu)\mathcal{T}^{0z}i\s_y \Omega & \frac{\kappa}{2}\vq^2 - \frac{\lambda}{2\nu} + J^2 D - \frac{i\s_y}{2\nu m^*}\Omega
    \end{matrix}\right), \label{K_matrix}
\end{equation}
\end{widetext}
and the four-component fluctuation vectors
\begin{equation}
M(q) = [\tvm_1^x(q) ~ \tvm_1^y(q) ~ \tvm_2^x(q) ~ \tvm_2^y(q)]^T
\end{equation}
and
\begin{align}\label{Jcurrent}
    \mathcal{J}(q) = \int\frac{\rmd^3 k}{(2\pi)^3} \begin{pmatrix}
        \J~\Psi^\dagger(k)\s_x \Psi(k-q)\\
        \J~\Psi^\dagger(k)\s_y \Psi(k-q)\\
        \J~\Psi^\dagger(k)\s_x \Psi(k-q)\\
        \J~\Psi^\dagger(k)\s_y \Psi(k-q)
    \end{pmatrix}.
\end{align}
The functions $D$, $T$ and $m^*$ are defined in the Appendix. We have also used the notation $q = (\Omega,\vq)$ and $k = (\omega, \vk)$ for bosonic and fermionic fields respectively. Performing the functional integral, we get the additional contribution to the Dirac action,
\begin{equation}
    \delta S_\rm{TI} = -\frac{1}{4}\int \frac{\rmd^3 q}{(2\pi)^3} \mathcal{J}^T(q) \mathcal{K}^{-1}(q) \mathcal{J}(-q).\label{S_TI}
\end{equation}
After calculating $\mathcal{K}^{-1}$, details of which are given in the Appendix, the effective action in the FM and AFM cases can be calculated separately. However, the resulting magnon-mediated interaction between the $\Psi$ fermions is given in the chirality basis rather than the spin basis. Therefore, we will derive the corresponding operator transformations for the $\Psi$ operators entering Eq.~(\ref{S_TI}) through the current vector $\mathcal{J}$ in Eq.~(\ref{Jcurrent}).

\subsection{Diagonalization of TI Hamiltonian}
The operator transformations are derived by diagonalizing the TI Hamiltonian including the interaction with the mean field magnetizations in Eq.~(\ref{L_MF}),
\begin{align}
    H_\rm{TI} &= \int \rmd^2 r ~ \Psi^\dagger [i\vf(\s_y\del_x - \s_x\del_y) - E_0\nabla^2\nonumber\\
    &\qquad\qquad\qquad- \mu- \J(1+\nu)\barm\s_z]\Psi,\nonumber\\
    &= \int \rmd^2 r ~ \Psi^\dagger \mathcal{H}_\rm{TI}\Psi.
\end{align}
Fourier transforming the Hamiltonian, and solving the eigenvalueproblem $\mathcal{H}_\rm{TI} \Psi_{\pm} = E \Psi_{\pm}$, we find the eigenenergies
\begin{equation}
    E_\pm(\vk) = E_0\vk^2 \pm \sqrt{\J^2\barm^2(1+\nu)^2 + \vf^2\vk^2}-\mu,\label{E}
\end{equation}
and eigenvectors
\begin{equation}
    \Psi_\pm(\vk) = \left(\begin{matrix}
        \psi_+(\vk)\\
        \psi_-(\vk)
    \end{matrix}\right) = \frac{1}{\sqrt{N_\vk}}\left(\begin{matrix}
        s_\vk^* & r_\vk\\
        -r_\vk & s_\vk
    \end{matrix}\right)\Psi(\vk)
    ,\label{op_transf}
\end{equation}
where we have defined the functions
\begin{subequations}\label{op_transf_func}
\begin{align}
    s_\vk &= \vf(k_y + i k_x),\\
    r_\vk &= \bar{J}\barm(1+\nu) + \sqrt{\bar{J}^2\barm^2(1+\nu)^2+\vf^2|\vk|^2},\\
    N_\vk &= r_\vk^2+|s_\vk|^2.
\end{align}
\end{subequations}
The subscripts $+$ and $-$ denote Dirac fermions with positive and negative chirality respectively. Note that the eigenvectors of $\mathcal{H}_\rm{TI}$ are unaffected by the value of $E_0$ since this is a diagonal term. If $\mu >0$, the conduction band will consist of $\psi_+$ fermions. Considering only the fermions which are free to interact, \ie projecting onto the conduction band, allows us to make the substitutions
\begin{equation}
    \psiu(k) \rightarrow \frac{s_\vk}{\sqrt{N_k}}\psi_+(k), ~\text{ and }~
    \psid(k) \rightarrow \frac{r_\vk}{\sqrt{N_k}}\psi_+(k), \label{cond_band_subst}
\end{equation}
in the effective action $\delta S_\rm{TI}$. This results in a momentum-dependent scattering form factor $\Lambda_{\vk\vk'}(\vq)$ characterizing the interaction between the fermions in the effective action, which we write as
\begin{equation}
\begin{aligned}
    \delta S_\rm{TI} &= \int \frac{\rmd^3 q}{(2\pi)^3}\int \frac{\rmd^3 k}{(2\pi)^3}\int \frac{\rmd^3 k'}{(2\pi)^3}\\
    &\quad\times  V_{kk'}(q)~\psi^\dagger(k+q)\psi^\dagger(k'-q)\psi(k')\psi(k). \label{S_TI_combined}
\end{aligned}
\end{equation}
with the interaction matrix defined as, 
\begin{equation}
    V_{kk'}(q) = -\J^2 D(q)\Lambda_{\vk\vk'}(\vq),
\end{equation}
where $D(q)$ is the magnon propagator. We refer to the Appendix for further details. If the effective action leads to an attractive interaction, it can be shown that this results in a superconducting instability, \eg by performing a mean field treatment of the effective theory. We will, however, not perform such an analysis, but rather focus on the type of effective interaction that arises  due to proximity of the magnetic layer. In the following two sections, we will analyse the effective action in the FM and AFM cases separately.

\section{Ferromagnetic case}\label{sec:FM}
In the ferromagnetic case, the magnon propagator is given by
\begin{equation}
    D^\rm{FM}(q) = \frac{\frac{\kappa}{2}\vq^2-\frac{J}{2a^2\barm}\Theta(1-\tau)}{\left(\frac{\Omega}{2m}\right)^2+ \left(\frac{\kappa}{2}\vq^2-\frac{J}{2a^2\barm}\Theta(1-\tau)\right)^2},
\end{equation}
where we have used the definitions of $D_i^{\alpha\alpha}$ and $T^{\alpha\alpha}$ given in Ref.~\onlinecite{Rex2017}, and $m$ is defined in the Appendix. Here, $a$ is the lattice constant, introduced when using $\pi/a$ as a cut-off in diverging momentum-integrals,\cite{Rex2017} $\Theta(x)$ is the Heaviside step function, and $\tau = t^2/\J^2\barm^2$ is a dimensionless parameter signifying the strength of the coupling between $\chi_1$ and $\chi_2$ relative to the coupling between $\chi_i$ and the magnetic moments, see Eq.~(\ref{H_surf}). 
{Assuming that the Dirac fermions move at speeds higher than the ferromagnetic magnons, which certainly holds for small momentum transfers $|\vq|$, we set $\Omega$ to zero in the magnon propagator. This yields }
\begin{equation}
    D^\rm{FM}(0, |\vq|) =\frac{1}{\frac{\kappa}{2}\vq^2-\frac{J}{2a^2\barm}\Theta(1-\tau)}. 
\end{equation}
Notice that if $\tau > 1$, $D^\rm{FM}(0,|\vq|)$ is positive for any $\vq$. Because the coupling constants $D_i^{\alpha\beta}$ and $T^{\alpha\beta}$ are discontinuous at $\nu = \tau$,\cite{Rex2017} values of $\tau \approx 1$ are excluded from the analysis in the ferromagnetic case.

\begin{figure}
    \includegraphics[width=0.65\columnwidth]{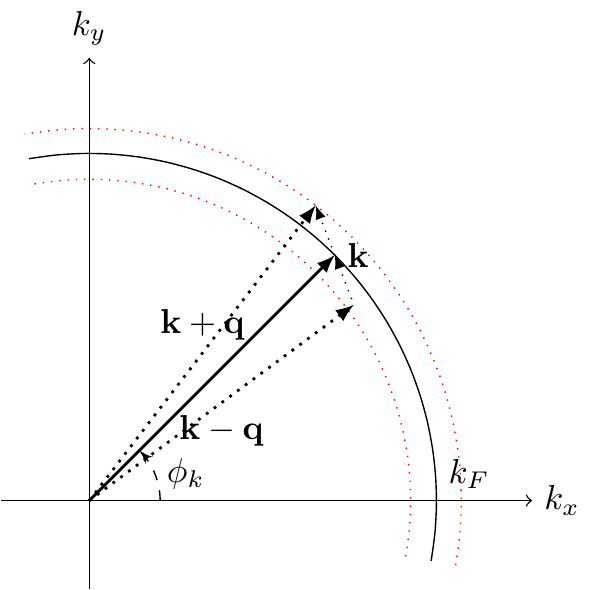}
    \caption{\label{fig:amperean_kinematics} The figure shows parts of the Fermi surface, and the momenta of the interacting particles, $\vk=\vk'$, $\vk + \vq$ and $\vk-\vq$. The figure illustrates that only small momentum transfers $|\vq|$ compared to $\kf$ are kinematically allowed, since the momenta must lie within a thin shell (red dotted lines) around $\kf$ (black line). This also implies that a process with momentum transfer $-\vq$ is necessarily kinematically allowed if the process with $\vq$ is allowed.}
\end{figure}

\begin{figure}
    \includegraphics[width=0.99\columnwidth]{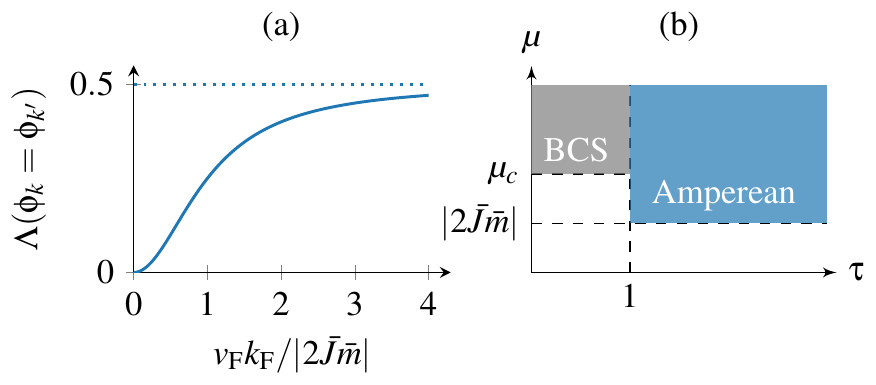}
    \caption{\label{fig:Lambda_FM} (a) Variation of $\Lambda(\phi_k=\phi_k')$ in Eq.~(\ref{Lambda_FM}). The form factor increases when increasing $\kf$, corresponding to moving the Fermi level away from the gap in the dispersion. $\Lambda(\phi_k=\phi_k')$ approaches $1/2$ in the limit $\vf \kf \gg |\J \barm|$. (b) Diagram showing the region in parameter space where Amperean and BCS pairing is possible for TI surface states coupled to a ferromagnet. $\mu = |2\J \barm|$ corresponds to $\kf = 0$.}
\end{figure}

Kargarian \etal \cite{Kargarian2016} found attractive interactions between particles with parallel momenta, dubbed Amperean pairing,\cite{Lee2007} in the high-doping regime. We expand this analysis to also include the gap in the Dirac fermion dispersion, \ie by not setting $\barm = 0$ in the operator transformations, Eqs.~(\ref{op_transf}) and (\ref{op_transf_func}). Since $\vk \approx \vk'$ for Amperean pairing, a process is only possible if both $\vk+\vq$ and $\vk-\vq$ lie within a thin shell of the Fermi level. This restricts the kinematically allowed values of $\vq$ to those with small $|\vq|$, as illustrated in Fig.~\ref{fig:amperean_kinematics}. Moreover, if a process with momentum transfer $\vq$ is possible, the process with momentum  transfer $-\vq$ is necessarily also possible. Thus, any term linear in $\vq$ in the form factor disappears when performing the $\vq$ integration in $\delta S_\rm{TI}$. 
Expanding the form factor in $\vf|\vq|/|\J\barm|$ and neglecting linear terms in $\vq$, we get to leading order
\begin{equation}
\Lambda(\phi_k,\phi_{k'}) = \frac{\vf^2\kf^2\left(2\J \barm + \sqrt{(2\J\barm)^2+\vf^2\kf^2}\right)^2\cos(\phi_k - \phi_{k'})}{2\left[\vf^2\kf^2 + (2\J \barm)^2 + 2\J \barm\sqrt{(2\J\barm)^2+\vf^2 \kf^2}\right]^2},\label{Lambda_FM}
\end{equation}
where we have set $|\vk|=|\vk'|=\kf$ and introduced the polar angle $\phi_k$ of each momentum in the $xy$ plane. Setting $\barm=0$, we get $\Lambda = \cos(\phi_k-\phi_{k'})/2$, which is in agreement with Ref.~\onlinecite{Kargarian2016}. This corresponds to the limit $\vf \kf \gg |\J\barm|$, for which the interaction is strongest. The interaction strength decreases for decreasing $\kf$, and disappears at $\kf = 0$, as illustrated in Fig.~\ref{fig:Lambda_FM}(a) This is as expected, since there must be a Fermi surface in the conduction band in order for interactions to be possible.

From the above results of the form factor and integrated magnon propagator, we see that the overall interaction matrix $V_{kk'}$ is negative for all kinematically allowed $\vq$ if $\tau > 1$, and $\vk$ and $\vk'$ are parallel. Hence, superconductivity with Amperean pairing is possible if $\tau > 1$. This is in agreement with the results in Ref.~\onlinecite{Kargarian2016}, which treats an analogous situation. Setting $\tvm_1 = \tvm_2$ in the magnon Lagrangian, Eq.~(\ref{L_mag_eff}), we notice that the ferromagnetic coupling between magnons,
\begin{equation}
    \tilde{\lambda}_1 + \tilde{\lambda}_2 - \lambda_\rm{eff} = -2J^2(D_1^{00}-D_1^{zz} - T^{00} + T^{zz}) =\frac{J}{a^2\barm}\Theta(1-\tau),
\end{equation}
disappears when $\tau > 1$, which is, again, similar to the situation discussed in Ref.~\onlinecite{Kargarian2016}.  

For $\tau<1$, $D^\rm{FM}(0,|\vq|)$ is negative for $|\vq| < \sqrt{J/a^2\barm\kappa}$, resulting in repulsive interactions, and changes sign as $|\vq|$ is increased. Since Amperean pairing is kinematically possible only  for small $|\vq|$, Amperean pairing is suppressed for increasing $J/\barm\kappa$. However, notice that for small $|\vq|$ and $\phi_k - \phi_{k'} \approx \pi$, corresponding to normal BCS pairing, the interaction matrix is attractive. 
Therefore, the possibility of BCS pairing is investigated further.

\begin{figure}
    \includegraphics{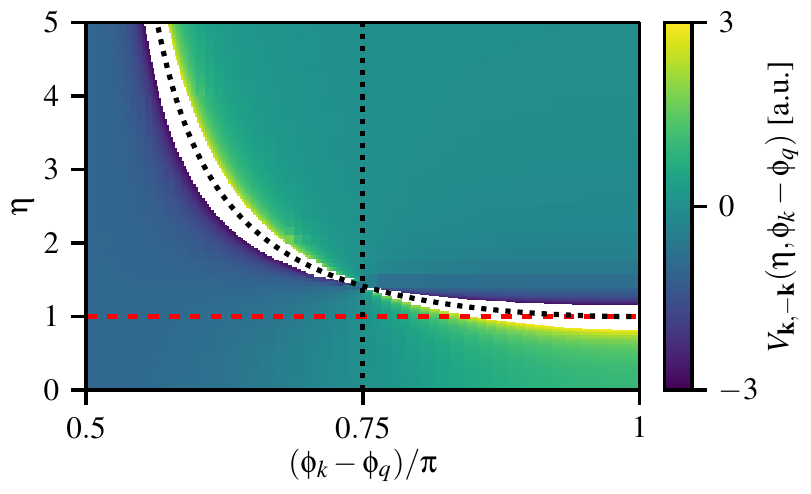}
    \caption{\label{fig:V_sign_FM} Plot of $V_{\vk,-\vk}$ in Eq.~(\ref{V_sign_FM}) as a function of $\eta$ and $\phi_k-\phi_q$, showing that the BCS pairing is attractive only for certain scattering angles $\phi_q$. The black dotted lines show where the interaction changes sign. For $\eta > 1$, shown by the red dashed line, integrating over the scattering angle gives a dominantly attractive pairing.}
\end{figure}

In the BCS case, $\vk' = -\vk$, the length of $\vq$ is less restricted, since $|\vk'-\vq|=|\vk+\vq| \approx \kf$ is satisfied for the same momentum transfer $\vq$. Requiring $|\vk| = |\vk+\vq| = \kf$, we find  
\begin{equation}
    |\vq| = \begin{cases} 
        -2\kf \cos(\phi_k-\phi_q), ~ \text{if } \pi \ge |\phi_k-\phi_q| \ge \frac{\pi}{2}\\
        0, \quad \text{otherwise}.
    \end{cases}
    \label{q_F}
\end{equation}
Inserted into the form factor, we find
\begin{equation}
     \Lambda^\rm{BCS}(\phi_k, \phi_q) = \frac{\vf^2 \kf^2 e^{2i(\phi_k-\phi_q)}\left(2\J\barm + \sqrt{\vf^2 \kf^2+(2\J\barm)^2}\right)^2}{2\left[\vf^2 \kf^2 + (2\J\barm)^2 + 2\J\barm\sqrt{\vf^2 \kf^2+(2\J\barm)^2}\right]^2}.\label{BCS_formfactor_FM}
\end{equation} 
Since the signs of $D^\rm{FM}(-2\kf \cos(\phi_k-\phi_q))$ and $\Lambda^\rm{BCS}(\phi_k,\phi_q)$ both vary with $\phi_k-\phi_q$, the overall sign of the real part of the interaction matrix will depend on $\phi_k-\phi_q$ as
\begin{equation}
     V_{\vk,-\vk} \propto -\frac{2a^2\barm}{J}\frac{\cos 2(\phi_k-\phi_q)}{\eta^2\cos^2(\phi_k-\phi_q)-1}, \text{if } \pi \ge |\phi_k-\phi_q| \ge \frac{\pi}{2}, \label{V_sign_FM}
\end{equation}
where $\eta \equiv 2\kf/\sqrt{J/a^2\barm\kappa}$. This quantity is plotted in Fig.~\ref{fig:V_sign_FM}, where it is clear that a BCS type interaction is both attractive and repulsive depending on the scattering angle.
Integrating $V_{\vk,-\vk}$ over $\phi_q$ gives a measure to whether most scattering angles are attractive or repulsive, and in this way gives a conservative estimate of when BCS pairing is possible. The results show that the overall interaction is attractive whenever $\eta > 1$, \ie when $2k_F > \sqrt{J/a^2\barm\kappa}$, which corresponds to chemical potential $\mu > \mu_c$, where
\begin{equation}
    \mu_c = \frac{E_0 J}{4a^2\barm \kappa} + \sqrt{(2\J\barm)^2 + \frac{\vf^2J}{4a^2\barm\kappa}}.
\end{equation}
Hence BCS pairing is possible for $\tau < 1$ and $\mu > \mu_c$. The attractive pairing is most dominant close to $\mu_c$, and decreases for increasing chemical potential. It is however important to note that the phase space of the pairing is reduced since not all scattering angles give attractive interactions, and the overall pairing is thus weakened compared to a normal BCS pairing.

In summary, we find that for $\tau > 1$, which corresponds to a disappearing $\tvm^2$ term in the magnon Lagrangian, superconductivity with Amperean pairing occurs. {For $\tau < 1$ and $\mu > \mu_c$ we instead have BCS pairing.} The pairing strength decreases for decreasing $\kf$ in the Amperean case, vanishing when the Fermi level lies inside the mass gap, while the BCS pairing is strongest close to $\mu_c$. A simplified diagram showing for which parameter values Amperean and BCS pairing occur is presented in Fig.~\ref{fig:Lambda_FM}(b).

\section{Antiferromagnetic case}\label{sec:AFM}
\begin{figure}
    \includegraphics{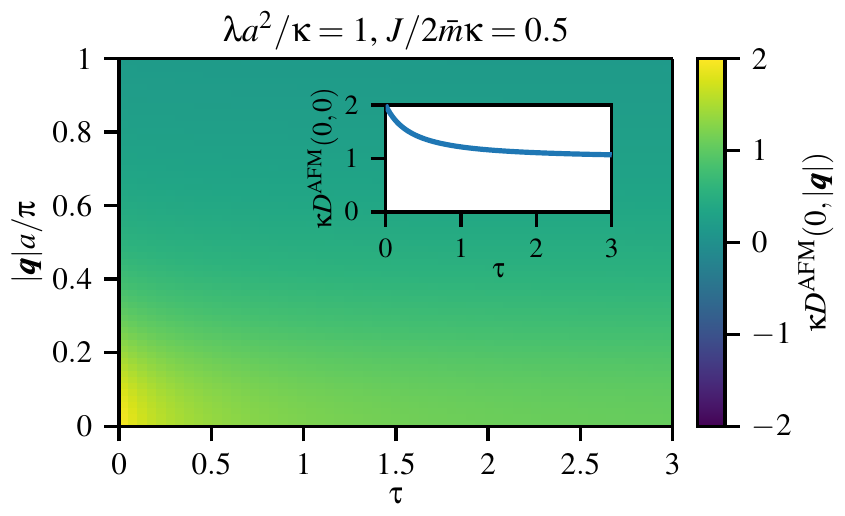}
    \includegraphics{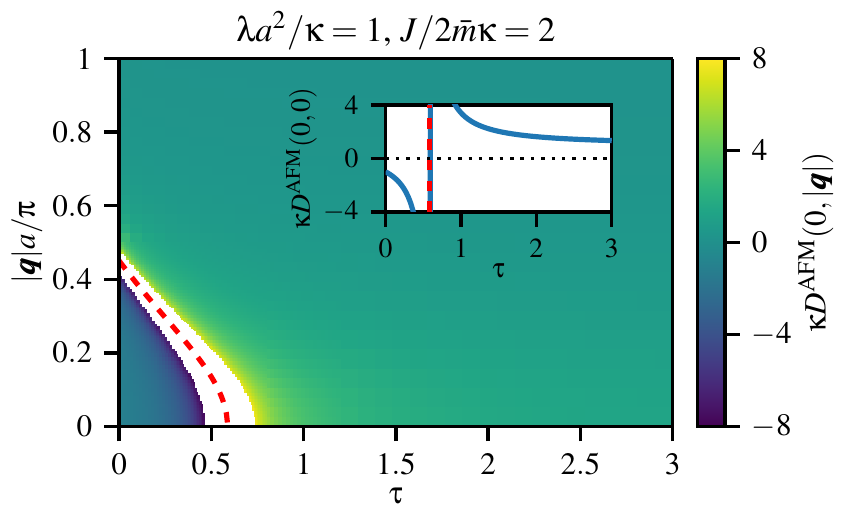}
    \caption{\label{fig:D_AFM} Plot of $D^\rm{AFM}$ as a function of $|\vq|$ and $\tau$ for $\lambda > J/2a^2\barm$ (top) and  $\lambda < J/2a^2\barm$ (bottom). In the former case $D^\rm{AFM}\ge 0$ for all $\vq$ and $\tau$. In the latter case however, the integrated propagator is negative in a region around $|\vq|=0$ and $\tau=0$. This region is bounded by the curve $q_c(\tau)$ (dashed red) given in Eq.~(\ref{q_c}), and increases for increasing $J/2a^2\barm\lambda$. {The white region indicates values outside the colorbar range. The propagator for $|\vq|=0$ is plotted in the insets.}}
\end{figure}

In the antiferromagnetic case, the net mean field magnetization is zero, and hence a gap is not opened in the dispersion. This also gives significantly simplified operator transformations, resulting in a scattering form factor
\begin{equation}
    \Lambda_{\vk\vk'}(\vq) = \frac{2\vk\cdot\vk' - i(\vk\times\vq - \vk'\times\vq)\cdot\hat{z}-\vk\cdot\vq + \vk'\cdot\vq}{4|\vk||\vk'|},
\end{equation}
where we have used $|\vk + \vq| \approx |\vk|$. The magnon propagator in the antiferromagnetic case is given by
\begin{equation}
\begin{aligned}
    &D^\rm{AFM}(q) =\\
     &\quad\frac{\frac{\kappa}{2}\vq^2-\frac{J}{2a^2\barm\sqrt{1+\tau}}}{\left(\frac{\Omega}{2m}\right)^2 + \left(\frac{\kappa}{2}\vq^2-\frac{J}{2a^2\barm\sqrt{1+\tau}}\right)\left(\frac{\kappa}{2}\vq^2-\frac{J}{2a^2\barm(1+\tau)^{3/2}} + \lambda\right)}.
\end{aligned}
\end{equation}
{The frequency of antiferromagnetic magnons typically lie in the microwave range,\cite{Kittel1963} and can therefore also be considered slow compared to the TI fermions, which have group velocities $\vf\sim 10^5$ m/s (see \eg Ref.~\onlinecite{Zhang2009}). Setting $\Omega = 0$ in the above propagator yields}
\begin{equation}
	D^\rm{AFM}(0,|\vq|) =\frac{1}{\frac{\kappa}{2}\vq^2-\frac{J}{2a^2\barm(1+\tau)^{3/2}} + \lambda}.
\end{equation}
Plots of $D^\rm{AFM}$ as a function of $|\vq|$ and $\tau$ for $\lambda > J/2a^2\barm$ and  $\lambda < J/2a^2\barm$ are shown in Fig.~\ref{fig:D_AFM}.
{From the figure we see that the propagator is positive for all $|\vq|$ and $\tau$ when $\lambda > J/2a^2\barm$. For $\lambda < J/2a^2\barm$ the propagator is positive for all $|\vq|$ if $\tau > \tau_c$, where
\begin{equation}
    \tau_\rm{c} = \left(\frac{|J/2a^2\barm|}{\lambda}\right)^{2/3} -1, \label{tau_c}
\end{equation}
and for $|\vq| > q_c$ if $\tau < \tau_c$, where
\begin{equation}
    q_c = \sqrt{\left|\frac{J}{2a^2\barm \kappa}\right|\frac{2}{(1+\tau)^{3/2}} - 2\frac{\lambda}{\kappa}}, \quad \tau < \tau_\rm{c}.\label{q_c}
\end{equation}
In the Amperean case we are again restricted to small momentum transfers, which to lowest order gives the form factor $\Lambda^\rm{Amp} = 1/2$. Hence magnon-induced Amperean pairing between Dirac fermions is possible either when $\lambda > J/2a^2\barm$, or when $\lambda < J/2a^2\barm$ and $\tau>\tau_c$.}

\begin{figure}
    \includegraphics[width=0.78\columnwidth]{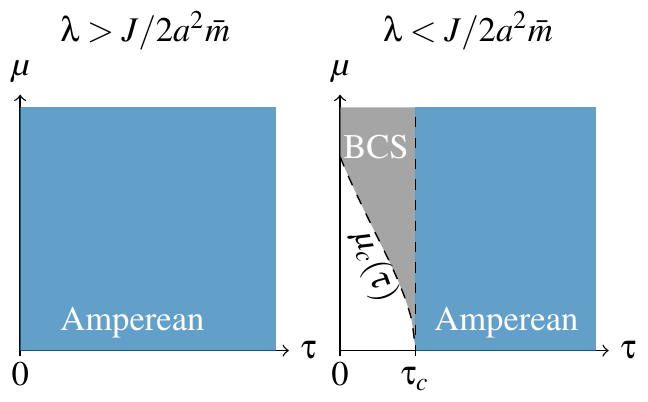}
    \caption{\label{fig:AFM_phasediag}Diagram showing the regions in parameter space where BCS and Amperean pairing is possible for TI surface states coupled to an antiferromagnet. BCS pairing is possible only when $\lambda < J/2a^2\barm$ when $\tau<\tau_c$ and $\mu > \mu_c(\tau)$.}
\end{figure}

For BCS pairing, however, we get the form factor $\Lambda^\rm{BCS} = e^{2i(\phi_k-\phi_q)}/2$, which corresponds to setting $\barm = 0$ in Eq.~(\ref{BCS_formfactor_FM}). The real part of the overall interaction can then be written
\begin{equation}
    V_{\vk,-\vk} \propto - \Big(\frac{J}{2a^2\barm(1+\tau)^{3/2}} - \lambda\Big)^{-1}\frac{\cos 2(\phi_k-\phi_q)}{\eta^2\cos^2(\phi_k-\phi_q)-1},
\end{equation}
for $\pi/2<|\phi_k-\phi_q|<\pi$. Here we have used Eq.~(\ref{q_F}), and defined $\eta = 2\kf/\sqrt{J/a^2\barm(1+\tau)^{3/2}\kappa - 2\lambda/\kappa}$. Again, the sign of the interaction depends on the parameter $\eta$ and the scattering angle $\phi_k-\phi_q$ in exactly the same way as in the FM case. Therefore the interaction is dominantly attractive when $\eta >1$, which corresponds to $\mu > \mu_c(\tau)$, where $\mu_c(\tau) = E_0 q_c(\tau)^2/4 + \vf q_c(\tau)/2$. Hence, BCS pairing can be realized when $\lambda < J/2a^2\barm$, $\tau <\tau_c$ and  $\mu>\mu_c$. This is a conservative limit, as there are attractive regions of phase space also when $\mu<\mu_c$.
The type of pairing realized for different values of $\tau$ and $\mu$ is shown in Fig.~\ref{fig:AFM_phasediag}.

$\tau_c$ corresponds to the value where $\tilde{\lambda}_1 + \tilde{\lambda}_2 - \lambda_\rm{eff}$, see Eq.~(\ref{ex_coupling_fluctuations}), changes sign from positive to negative, \ie the point where the ferromagnetic coupling between spins on each of the two sublattices becomes weaker than the antiferromagnetic coupling between spins on different sublattices. {Thus, for both the FM and AFM cases, BCS pairing seems to be possible when the quadratic $\tvm^2$ term dominates over the interlattice coupling.}

\section{Summary}\label{sec:conclusion}
We have studied the possible electron pairing due to magnetic fluctuations at the interface of a TI and a FM or AFM insulator.
In the FM case, we have expanded the results of Ref.~\onlinecite{Kargarian2016} to be valid also for chemical potentials close to the gap in the TI fermion dispersion. We find that for $\tau > 1$, which corresponds to a vanishing quadratic term $\propto\tvm^2$ in the magnon Lagrangian, Amperean pairing occurs. The pairing strength decreases for decreasing $\kf$ and vanishes when the chemical potential lies inside the mass gap. For $\tau < 1$, Amperean pairing is suppressed for increasing $J/\barm\kappa${, and instead BCS pairing occurs above a critical chemical potential.
{In the AFM case BCS pairing is realized only when the ferromagnetic coupling between magnons on the same sublattice exceeds the antiferromagnetic coupling between magnons on different sublattices. For other parameter values, Amperean pairing is realized with an interaction strength indepentent of the chemical potential.} In both the FM and AFM case, the BCS pairing has a limited phase space compared to the regular BCS interaction, and could therefore be a weak effect, depending on the chemical potential of the system.

In conclusion, magnetic fluctuations at the interface between a TI and magnetic insulator can mediate attractive interactions between Dirac fermions, giving pairing of both BCS and Amperean type, depending on the degree of anisotropy of the magnetic fluctuations in the system. Investigating other magnetic configurations, such as ferrimagnetic insulators, would be an interesting further development. 
We also leave it for future work to consider bilayers involving magnetic metals, where a similar pairing mechanism is likely to remain in effect. For the metallic FM case, non-$s$-wave pairing has already been reported in recent experiments on superconducting Ni-Bi bilayers \cite{Gong2015, Gong2017, Wang2017}.

\begin{acknowledgments}
H.G.H. and A.S. were supported by the Research Council of Norway through Grant Number 250985, "Fundamentals of Low-dissipative Topological Matter", and Center of Excellence Grant Number 262633, Center for Quantum Spintronics. S. R. was supported by DFG through SPP 1666, "Topological Insulators: Materials - Fundamental Properties - Devices".
\end{acknowledgments}

\appendix*

\section{Calculation of effective TI action}\label{app:calcSeff}
In the cases $\barm_2/\barm_1 = \nu = \pm 1$, we have the relations $D_1^{00} = D_2^{00}$, $D_1^{zz} = D_2^{zz}$,  $\mathcal{D}_1^{0z} = \nu \mathcal{D}_2^{0z}$, while $T^{zz}$ and $\mathcal{T}^{0z}$ are zero in the antiferromagnetic case. To handle this, we write $T^{zz} = (1+\nu)T^{zz}/2$ and $\mathcal{T}^{0z} = (1+\nu)\mathcal{T}^{0z}/2$. Inserting this into the magnon action, and rewriting in vector form, we defined the matrix $\mathcal{K}$ in Eq.~(\ref{K_matrix}) using the functions
\begin{subequations}
\begin{align}
    D &= D_1^{00} - D_1^{zz},\\
    T &= T^{00} - \frac{1+\nu}{2}T^{zz},\\
    \frac{1}{2m^*}  &= \frac{1}{2\barm} - 2J^2\mathcal{D}_1^{0z},
\end{align}
\end{subequations}
for notational simplicity.

The inverse of $\mathcal{K}$ can be written on the form
\begin{equation}
    \mathcal{K}^{-1} = \frac{1}{\det \mathcal{K}} \begin{pmatrix}
        A_0 + A_y i\s_y & B_0 + B_y i\s_y\\
        B_0 + B_y i\s_y & A_0 + \nu A_y i\s_y,
    \end{pmatrix}
\end{equation}
where
\begin{widetext}
\begin{equation}
\begin{aligned}
    \det \mathcal{K} = {} &
    \left(\frac{\kappa}{2}\vq^2+J^2 D\right)^2\Bigg\{\left[\left(\frac{\kappa}{2}\vq^2+J^2 D\right) - \nu\lambda \right]^2 +2\left[\left(\frac{\Omega}{2m^*}\right)^2 + (1+\nu)^2(\mathcal{T}^{0z}\Omega)^2\right] - 2\lambda T - 2T^2\Bigg\}\\
    &+2\left(\frac{\kappa}{2}\vq^2+J^2 D\right)\Bigg[\nu\lambda T(\lambda + T)+4T(1+\nu)\frac{\mathcal{T}^{0z}\Omega^2}{2m^*} - \lambda\nu \left(\frac{\Omega}{2m^*} - (1+\nu)\mathcal{T}^{0z}\Omega\right)^2\Bigg]\\
    &+ T^2\Bigg[(T +\lambda)^2 + 2\nu\left(\frac{\Omega}{2m^*}\right)^2 + 2(1+\nu)^2(\mathcal{T}^{0z}\Omega)^2\Bigg] + (1+\nu)^2\frac{\lambda^2}{2^2}\Bigg(\frac{\Omega}{2m^*}-2\mathcal{T}^{0z}\Omega\Bigg)^2 \\
    &+ 2\lambda\nu T\left(\frac{\Omega}{2m^*}  - (1+\nu)\mathcal{T}^{0z}\Omega\right)^2 + \Bigg[\left(\frac{\Omega}{2m^*}\right)^2 - (1+\nu)^2(\mathcal{T}^{0z}\Omega)^2\Bigg]^2,
\end{aligned}
\end{equation}
and 
\begin{subequations}
\begin{align}
    A_0 = {} &
    \left(\frac{\kappa}{2}\vq^2+J^2 D\right)^3 - \frac{3\nu\lambda}{2}\left(\frac{\kappa}{2}\vq^2+J^2 D\right)^2 + \left(\frac{\kappa}{2}\vq^2+J^2 D\right)\Bigg[\frac{\lambda^2}{2} - \lambda T - T^2 + \left(\frac{\Omega}{2m^*}\right)^2+ (1+\nu)^2(\mathcal{T}^{0z}\Omega)^2\Bigg]\nonumber\\
    &+ \frac{\nu}{2}\lambda T(\lambda + T) + 2(1+\nu)T \frac{\mathcal{T}^{0z}\Omega^2}{2m^*} - \frac{\nu\lambda}{2}\left(\frac{\Omega}{2m^*} - (1+\nu)\mathcal{T}^{0z}\Omega\right)^2,\label{A0}\\
    A_y = {} & 
    \left(\frac{\kappa}{2}\vq^2+J^2 D\right)^2\frac{\Omega}{2m^*} + \left(\frac{\kappa}{2}\vq^2+J^2 D\right) \Bigg[(1+\nu)\left(2T+\lambda\right) \mathcal{T}^{0z}\Omega - \nu\lambda\frac{\Omega}{2m^*}\Bigg]\nonumber\\
    &+\frac{\Omega}{2m^*}\Bigg[\left(\frac{\Omega}{2m^*}\right)^2 - (1+\nu)^2(\mathcal{T}^{0z}\Omega)^2+\nu T^2 + (1+\nu)\frac{\lambda^2}{4}+\nu \lambda T\Bigg] - (1+\nu) \frac{\lambda}{2}(\lambda + 2T) \mathcal{T}^{0z}\Omega,\\
    B_0 = {} & -\frac{1}{2}\left(\frac{\kappa}{2}\vq^2+J^2 D\right)^2(\lambda + 2T) + \left(\frac{\kappa}{2}\vq^2+J^2 D\right)\left[\frac{\lambda}{2}(\lambda + 2T) + (1+\nu)^2\frac{\mathcal{T}^{0z}\Omega^2}{2m^*}\right] + \frac{1}{2}\lambda^2 T\nonumber\\
    &+ \nu T\left[\left(\frac{\Omega}{2m^*}\right)^2 + (1+\nu)^2(\mathcal{T}^{0z}\Omega)^2\right] + T^2\left(T+\frac{3}{2}\lambda\right) + \nu \frac{\lambda}{2}\left[\frac{\Omega}{2m^*} - (1+\nu)\mathcal{T}^{0z}\Omega\right]^2,\label{B0}\\
    B_y = {} & -\frac{\Omega}{2m^*}(1+\nu)\left(\frac{\lambda}{2} + T\right)\left(\frac{\kappa}{2}\vq^2+J^2 D - \frac{\lambda}{2}\right)\nonumber\\
    &- (1+\nu)\mathcal{T}^{0z}\Omega\left[(1+\nu)^2(\mathcal{T}^{0z}\Omega)^2 + \left(\frac{\kappa}{2}\vq^2+J^2 D - \frac{\lambda}{2}\right)^2 + \left(\frac{\lambda}{2} + T\right)^2 - \left(\frac{\Omega}{2m^*}\right)^2\right].
\end{align}
\end{subequations}
The above equations have been simplified using $\nu^2 =1$ and $1/\nu = \nu$, and are therefore valid only when $\barm_2 = \pm \barm_1$.

Performing the matrix multiplication in Eq.~(\ref{S_TI}) using the above form of $\mathcal{K}^{-1}$ and the definition of $\mathcal{J}(q)$ in Eq.~(\ref{Jcurrent}), we get
\begin{equation}
\begin{aligned}
    \delta S_\rm{TI} &= -\J^2\int \frac{\rmd^3 q}{(2\pi)^3}\int \frac{\rmd^3 k}{(2\pi)^3}\int \frac{\rmd^3 k'}{(2\pi)^3}
    \Bigg\{\frac{A_0 + B_0}{\det \mathcal{K}} \big[\psiu^\dagger(k)\psid^\dagger(k')\psiu(k'+q)\psid(k-q) 
    + \psid^\dagger(k)\psiu^\dagger(k')\psid(k'+q)\psiu(k-q)\big]\\
    &\qquad\quad+ i\frac{A_y(1+\nu) + 2B_y}{2\det \mathcal{K}} \big[\psiu^\dagger(k)\psid^\dagger(k')\psiu(k'+q)\psid(k-q) 
    - \psid^\dagger(k)\psiu^\dagger(k')\psid(k'+q)\psiu(k-q)\big] \Bigg\}.
\end{aligned}
\end{equation}
\end{widetext}
In the antiferromagnetic case, $A_y(1+\nu) + 2B_y$ is exactly equal to zero. In the ferromagnetic case however, this term has an overall factor of $\Omega$, making it less divergent in the low-frequency limit. We will therefore neglect this term.\cite{Kargarian2016}

Projecting onto the conduction band using Eq.~(\ref{cond_band_subst}), we get the effective action
\begin{equation}
\begin{aligned}
    \delta S_\rm{TI} &= -\J^2\int \frac{\rmd^3 q}{(2\pi)^3}\int \frac{\rmd^3 k}{(2\pi)^3}\int \frac{\rmd^3 k'}{(2\pi)^3}\\
    &\times D(q)\Lambda_{\vk\vk'}(\vq)~\psi^\dagger(k+q)\psi^\dagger(k'-q)\psi(k')\psi(k),
\end{aligned}
\end{equation}
where we have dropped the subscript $+$ for notational simplicity, and defined the magnon propagator
\begin{equation}
    D(q) = \frac{A_0(q) + B_0(q)}{\det \mathcal{K}(q)},
\end{equation}
and the scattering form factor
\begin{equation}
    \Lambda_{\vk\vk'}(\vq) = \frac{s^*_{\vk+\vq} r_{\vk'-\vq} s_{\vk'} r_{\vk} + r_{\vk+\vq} s^*_{\vk'-\vq} r_{\vk'} s_{\vk}}{\sqrt{N_\vk N_{\vk'} N_{\vk-\vq} N_{\vk'+\vq}}}.
\end{equation}

Defining the parameter $m$ such that
\begin{equation}
     \frac{1}{2m} = \frac{1}{2\barm} - 2J^2\mathcal{D}_1^{0z} - J^2(1+\nu)\mathcal{T}^{0z},
\end{equation}
and using the results in Eqs.~(\ref{A0}) and (\ref{B0}), we get the ferromagnetic propagator ($\nu = 1$)
\begin{equation}
    D^\rm{FM}(q) = \frac{\frac{\kappa}{2}\vq^2+J^2 (D + T)}{\left(\frac{\Omega}{2m}\right)^2 + \left(\frac{\kappa}{2}\vq^2+J^2 (D+T)\right)^2}
\end{equation}
and the antiferromagnetic propagator ($\nu = -1$)
\begin{equation}
\begin{aligned}
    &D^\rm{AFM}(q) =\\
    &\quad \frac{\frac{\kappa}{2}\vq^2+J^2 (D - T)}{\left(\frac{\Omega}{2m}\right)^2 + \left(\frac{\kappa}{2}\vq^2+J^2 (D-T)\right)\left(\frac{\kappa}{2}\vq^2+J^2 (D+T) + \lambda\right)}.
\end{aligned}
\end{equation}

\end{document}